\RequirePackage{fix-cm}

\documentclass[smallextended]{svjour3}       
\usepackage[utf8]{inputenc}

\usepackage{booktabs}
\usepackage{pdflscape}
\usepackage{geometry}
\usepackage{setspace}
\usepackage{graphicx}
\usepackage{dcolumn}
\usepackage{graphicx}

\usepackage{amssymb,amsmath,amsthm}
\usepackage{ifthen}
\usepackage{setspace}
\usepackage{url}
\usepackage[usenames,dvipsnames]{color}
\usepackage{hyperref}
\usepackage[flushleft]{threeparttable}
\usepackage{rotating}
\usepackage{cite}
\usepackage{multirow}
\usepackage{quoting}
\usepackage{makebox}
\usepackage{caption}
\usepackage{subfig}
\usepackage{fancyhdr}
\usepackage{geometry}
\usepackage{bbm}
\usepackage{setspace}
\usepackage{amsmath}
\usepackage{cite}
\usepackage{rotating}
\usepackage{tabularx}
\usepackage{array}
\usepackage{microtype}
\usepackage{indentfirst}
\usepackage{booktabs}
\usepackage{float}
\usepackage{verbatim}
\usepackage[super]{nth}
\usepackage{titlesec}
\usepackage{color,soul}
\usepackage{natbib}
\smartqed  %
\usepackage{graphicx}
\AtBeginEnvironment{tabular}{\doublespacing}
\begin{document}

\title{Slingshot spiders build tensed underdamped webs for ultrafast launches and speedy halts

}


\author{Elio J. Challita* \and
        Symone Alexander* \thanks{*These authors contributed equally} \and
        Sarah Han \and
        Todd A. Blackledge \and
        Jonathan A. Coddington \and
        Sunghwan Jung \and
        M. Saad Bhamla
}


\institute{Elio J. Challita \at
             Chemical and Biomolecular Engineering, Georgia Institute of Technology, Atlanta, GA 30311, USA \\
             Mechanical Engineering, Georgia Institute of Technology, Atlanta, GA 30311, USA
           \and
           Symone Alexander \at
               Chemical and Biomolecular Engineering, Georgia Institute of Technology, Atlanta, GA 30311, USA 
               \and
           Sarah Han \at
           Department of Biology, Integrated Bioscience Program, The University of Akron, Akron, OH 44325 , USA
           \and
           Todd A. Blackledge \at
           Department of Biology, Integrated Bioscience Program, The University of Akron, Akron, OH 44325, USA
             \and
           Jonathan A. Coddington \at
           Smithsonian Institution, National Museum of Natural History, 10th and Constitution, NW Washington, DC 20560, USA
            \and
           Sunghwan Jung \at
           Department of Biological and Environmental Engineering, Cornell University, Ithaca, NY 14853, USA
            \and
           M. Saad Bhamla \at
              Chemical and Biomolecular Engineering, Georgia Institute of Technology, Atlanta, GA 30311, USA\\
              \email{saadb@chbe.gatech.edu}
}

\date{Received: date / Accepted: date}

\maketitle

\begin{abstract}
We develop a mathematical model to capture the web dynamics of slingshot spiders (Araneae: Theridiosomatidae), which utilize a tension line to deform their orb webs into conical springs to hunt flying insects. Slingshot spiders are characterized by their ultrafast launch speeds and accelerations (exceeding 1300 $m/s^2$), however a theoretical approach to characterize the underlying spatiotemporal web dynamics remains missing. To address this knowledge gap, we develop a 2D-coupled damped oscillator model of the web. Our  model reveals three key insights into the dynamics of slingshot motion. First, the tension line plays a dual role: enabling the spider to load elastic energy into the web for a quick launch (in milliseconds) to displacements of 10-15  body lengths, but also enabling the spider to halt quickly, attenuating inertial oscillations. Second, the dominant energy dissipation mechanism is viscous drag by the silk lines - acting as a low Reynolds number parachute. Third, the  web exhibits underdamped oscillatory dynamics through a finely-tuned balance between the radial line forces, the tension line force and viscous drag dissipation. Together, our work suggests that the conical geometry and tension-line enables the slingshot web to act as both an elastic spring and a shock absorber, for the multi-functional roles of risky predation and self-preservation. 

\end{abstract}

\section*{Introduction}
\label{intro}
The sit and wait strategy of many orb-weaving spiders is well known for the ability to effectively snare prey using sticky silk and the rapid dissipation of the captured prey's kinetic energy~\citep{Kelly2011-lc,Sensenig2012-id,Yu2015-yr,Das2017-ba}.
Slingshot spiders, orb-weavers known for extending and releasing their webs like slingshots to capture prey, evolved an orb web modified with a tension line attached at its central hub~\citep{Alexander2020-bu,Coddington1986-du,Hingston1932-bt,Wienskoski2010-bd,Alves_undated-eg, Eberhard1981-as, Eberhard1990-uq, Eberhard1986-kw, Coddington2005-yw}. The tension line enables the spider to deform  its  web into a 3-D conical structure, loading elastic energy into the radial lines that ultimately facilitates rapid accelerations ($>1300\; m/s^2$) for capturing flying prey (SI Movie 1), or possibly avoiding predation~\citep{Alves_undated-eg}. Although in previous work, we  described how the slingshot spider loads the web as a spring to achieve ultrafast motion \citep{Alexander2020-bu}, it remains unclear  how the spider leverages its unique web and tension line to decelerate quickly and come to a halt, either after successful capture or missing of prey~\citep{Eberhard1990-uq}, and to reset its web to potentially fire again. What are the relevant physical forces governing energy storage and energy dissipation in the web? How does the conical geometry affect the slingshot spatial and temporal dynamics? The role of the tension line in enabling the spider to load energy into the web as a spring is known, but what role does the tension play in stopping the spider's motion? To address these open questions, we develop a mathematical model of the slingshot spider web dynamics in this paper.

The slingshot spider grips the center of its web with four rear legs while incrementally pulling and twisting the tension line and coiling the silk with claws on its anterior legs and pedipalps (non-locomotor anterior appendages) to store elastic energy in the radial lines of the web
(Fig. 1a)~\citep{Alexander2020-bu,Coddington1986-du}. Upon sensing external stimuli (for example a finger snap), the spider releases the tension line, catapulting both the web and spider backwards (Fig. 1b). Multiple trajectories ($n = 5$) are provided in Fig. 1c, with the full displacement taking place in around 30 ms. The dominant movement is mostly in the y-direction (10 - 15 mm or 10-15 $\times$ body lengths) with smaller movements ($\pm 1.1$~mm, less than one body length) in the x-direction. Spiders achieve vertical speed of up to 4.2 m/s ($v_{max} = 4.16 \pm 0.07\;m/s$) and accelerations exceeding 1300$\;m/s^2$ ($a_{max} = 1163 \pm 144\;m/s^2$) ~\citep{Alexander2020-bu}. Despite this ultrafast millisecond motion, we observe that the resultant vibrational response of the web attenuates within milliseconds as well - suggesting that the web design facilitates both energy storage and speed as well as energy dissipation, which could potentially improve prey capture, reduce the probability of damage to spider or web, and/or facilitate rapid reset and reloading of the web.

\begin{figure}
 \centering
  \includegraphics[width=1\textwidth]{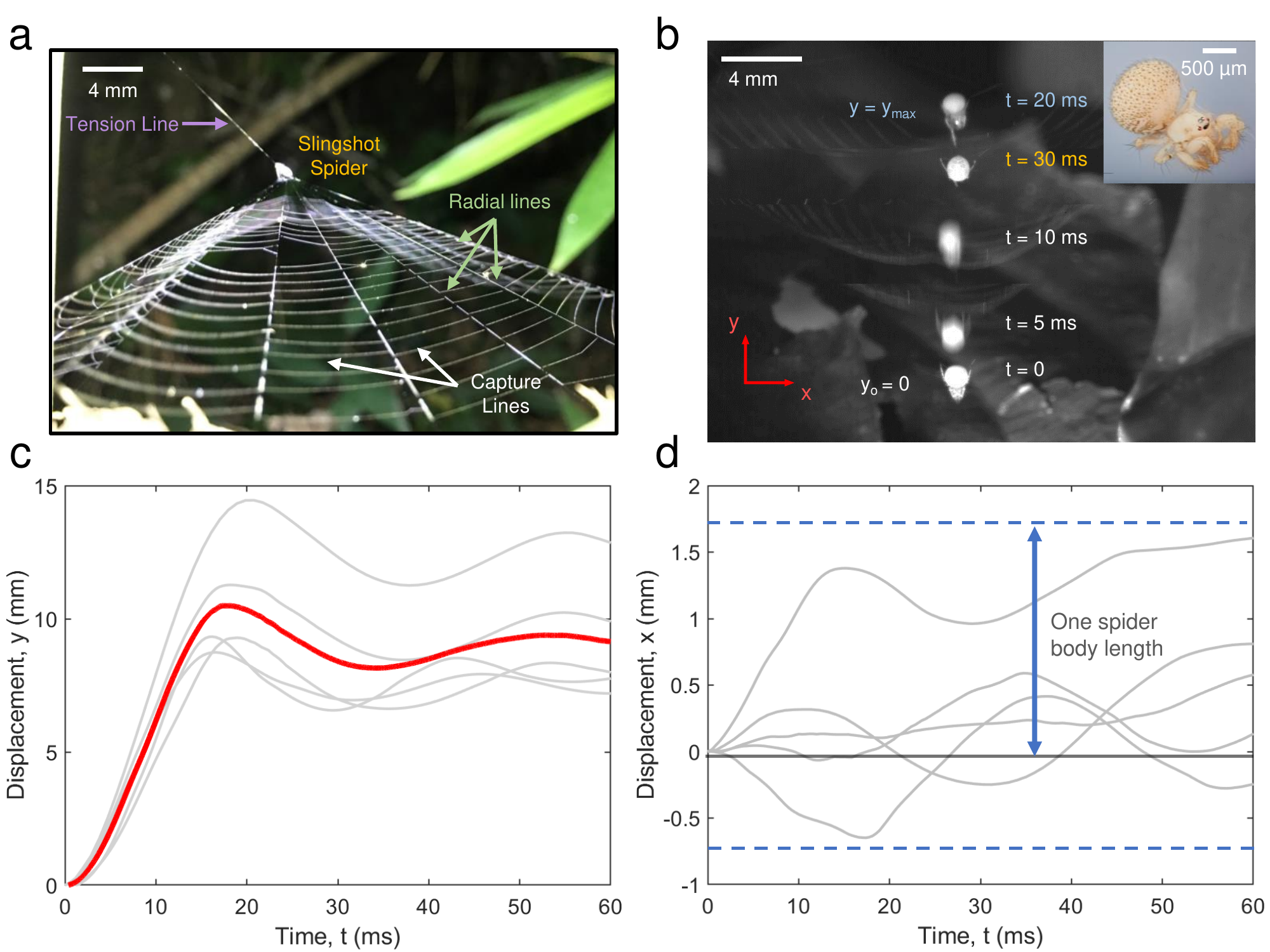}
\caption{\textbf{Slingshot spider web dynamics}. \textbf{a} The spider loads elastic energy by pulling on the tension line, thereby stretching the radial lines and deforming the web and forming a conical web structure. The spider remains in this position awaiting its prey. We note from field observations that the spider web does not seem to have a specific preference for orientation with respect to gravity.  \textbf{b} Upon sensing a stimulus the slingshot spider releases the web moving explosively, almost 9 times its body length in less than 30 ms. The web is oriented horizontally with the tension line at the bottom of the frame, so that the primary motion upon release in the positive $y$ direction facing upwards in this case. \textbf{c} Quantification of the multiple trajectories of the same single slingshot spider for repeated launches (triggered by snapping fingers). The red line denotes the average trajectory of the multiples displacement curves. We adjust trajectories so that they all start at $(x=0 , y=0)$ at $t=0$. \textbf{d} Movement in the $x$ direction is negligible before the first oscillation.}

\label{fig:1}      
\end{figure}

Motivated by the  rapid attenuation of the slingshot spiders motion as well as its non-planar configuration, we set to theoretically model the dynamics of the slingshot spider web. We treat the slingshot motion like the dynamic response of a step input typically used in process control design~\citep{Kuo1987-km}. Specifically, we focus on spatiotemporal parameters such as the rise time, overshoot and settling time. Modeling the forces in the radial lines and the tension line allows us to explore how they enable both rapid movement and quick return back to the equilibrium position.  The mathematical model described herein offers insights about the balance of elastic forces in slingshot spider webs and the opportunity to understand their design in ways that are difficult to achieve experimentally. 

\section*{Methods}
\subsection*{Field Videography}
Field work was conducted in Puerto Maldonado, Peru at the Tambopata Research Center ($-13.134^\circ$,  $-069.609^\circ$). Research permit no. 654-2018-GOREMAD-GRRNYGMA-DRFFS was obtained by the Gerencia Regional Forestal y de Fauna Silvestre. The spiders were located by scanning dead branches and leafy plants for their conical webs and then snapping fingers near to the web to confirm slingshot motion. A Chronos 1.4 high speed camera (Krontech) was utilized for high speed video recording (up to 38,500 fps) in conjunction with a field portable Zaila high intensity light with portable battery packs. Field Videos were captured at 1057 fps.  

\subsection*{Field specimen and silk collection}
The spider specimen was identified to the best of our ability as an  undescribed species in the genus \textit{Epeirotypus} sp. (Araneae: Theridiosomatidae) (see SI text for images of organism including epigynum). Videography was performed utilizing spiders that had built webs in their natural habitat. After observation and videography, spider specimens were collected and stored in 200 proof ethanol for species identification and further analysis. Silk samples were collected using a notched microscope slide.

\subsection*{High speed video analysis}
Matlab was employed to analyze the highspeed video obtained in the field. The code was written to identify the spider in each frame using an intensity threshold and record the location of its centroid. Accurate tracking was verified utilizing a binary output video highlighting the spider as white and setting the background as black. The code converted the units of the centroid measurements from pixels to meters and calculated elapsed time from the frame rate and number of frames to calculate displacement, velocity, and acceleration.

\section*{Mathematical Model}



We mathematically model the spatiotemporal dynamics of the motion of the slingshot spider as a 2D mass-spring model in $x$-$y$ direction (Fig. 2a). The model consists of three springs: two symmetric springs extended in the radial/horizontal $x$ direction ($F_{r1,2}$) and one in the vertical $y$ direction that represents the tension line ($F_t$). We model the spider as a point mass ($m_s$) located at the intersection of the three springs (Fig. 2). In the following subsections, we describe different components of the model in detail along with relevant assumptions and limitations.

\begin{figure}[ht]
 \includegraphics[width=1\textwidth]{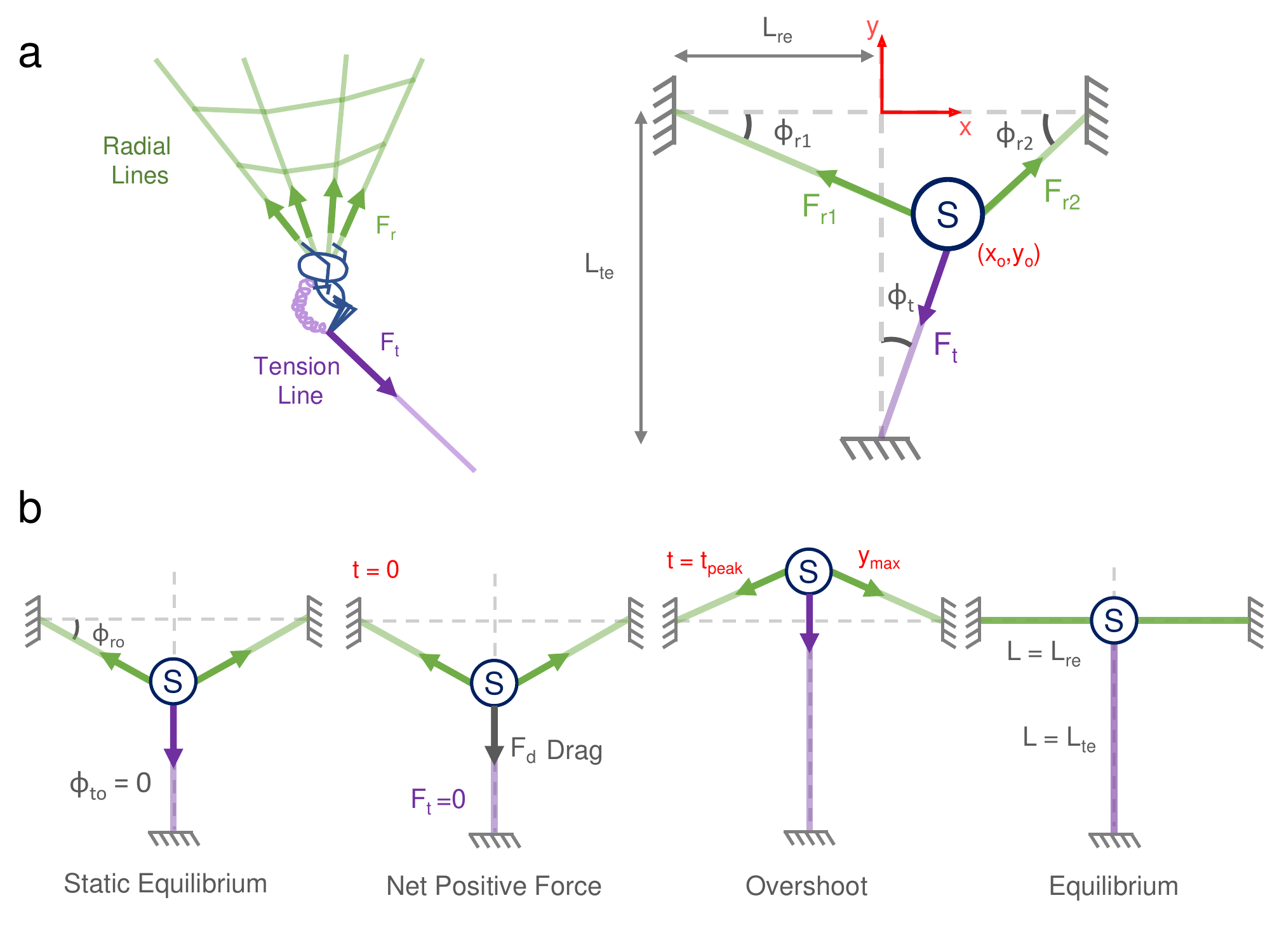}
\caption{ \textbf{Schematic of the mathematical model.} \textbf{a} We model the conical web structure as two symmetric horizontal radial springs and a vertical tension line located in a 2D plane. We assume that the slingshot spider is a point mass having a characteristic length $D_s$ on the order of $\sim 1.3\; mm$ located at the intersection of radial lines and the tension line.  We define $L_{r,eq}$ and $L_{t,eq}$ as the equilibrium length of the the radial lines and the tension line respectively.The spider starts at $(x_o,y_o)$ at $t=0$. Each force has a projected component in the $x$ and $y$ directions. Our model ignores the effects of gravity on the slingshot dynamics and the positive $y$ direction is set facing the upward direction.   \textbf{b} We consider the case of a pure 1D motion where at $t<0$, the radial lines forces are balanced by the tension line force. At $t=0$, the spider releases the tension  resulting in a net force in the positive (upward in this particular case) y-direction caused by the radial lines and opposed by the air drag. When the spider crosses the equilibrium point, the tension line starts stretching, resulting in a restoring downward force. The spider continues oscillating, while slowly approaching equilibrium. }
\label{fig:2}      
\end{figure}
\subsection*{2D web geometry}

We define $\phi_{r1}$ and $\phi_{r2}$ as the angles formed between the radial lines and the $x$-axis and $\phi_{t}$ as the angle between the tension line and the $y$-axis (Fig. 2b). We prescribe two equilibrium points to the model (Fig. 2b). The first equilibrium point is before the spider launches ($t<0$, $\phi_{r1,2}=\phi_{r0}$, Fig. 2c), when the spider has deformed the web into a cone and waits for a prey to come within striking distance. At this point the forces in the radial springs are balanced by the force in the tension line and the system is in a static equilibrium. The second equilibrium point is defined when the all the  web lines (radial and tension line) are at their equilibrium lengths ($L_r$=$L_{r,eq}$ , $L_t$=$L_{t,eq}$ , $\phi_{r1,2}$ =  $\phi_{t}$ = 0, Fig. 2c). This point occurs when the spider's motion has ceased and all the web forces are equal to zero (Fig. 2c). We note that in the actual slingshot spider web these equilibrium points may be shifted due to the asymmetric architecture of the web. However for the purposes of our model, this assumption is valid as we are assuming that the springs are linear and that the variations in the forces are more important to the dynamics compared to the absolute base forces themselves. In a sense, this is similar to the treatment of a vertical mass-spring system where the effects of gravity are ignored since the equilibrium point is defined when the weight is balanced by the initial extension of the spring.

\subsection*{Elastic silk springs}
 We make two assumptions to model the silk web lines (radial and tension lines) as elastic force-generating springs. First, the silk lines are approximated as linear Hookean springs ($F=K\cdot \Delta L$) with no internal viscous damping. We recognize that this assumption is a significant simplification since actual silk fibers behave as complex and nonlinear viscoelastic materials~\citep{Tietsch2016-vq,Yu2015-yr,Gosline1986-mi, Gosline1984-mn,Kelly2011-lc}.  The spring constant $K$ is defined by $K=EA/L_{eq}$ , where  $A$ is the cross-sectional area of the silk measured using SEM images (See SI text Fig. 4), and $L_{eq}$ is the equilibrium length and $E$ is the silk's young modulus.
The silk's young modulus ranges between $\sim 1- 10$ GPa and is obtained from the nonlinear stress-strain curves of the major ampullate (MA) silk of various spider species ~\citep{Gosline1999-jr,Ko2001EngineeringSilk} .
 
 Second, unlike conventional springs that can exert a push or pull force depending on compression or extension, our modeled silk springs react only to extension to exert a pulling force - they cannot push \citep{jung2014capillary}. Mathematically, we incorporate this through the Heaviside function $\Theta(\Delta L)$, which is defined as $\Theta(\Delta L)=1$ when the silk spring is extended ($\Delta L >0$) and $\Theta(\Delta L)=0$ for compression $\Delta L <0$. 
 

\subsection*{Energy dissipation to the environment through aerial drag}
The energy of the system is dissipated by viscous drag to the environment by the fast moving spider and web. We ignore any friction that might arise between the spider pedipalps. Viscous damping in the silk lines itself is also ignored as discussed earlier.
The aerodynamic drag of an object traveling through a fluid (air) depends on several factors such as its geometry, dimensions  and flow conditions. The Reynolds number, which determines the effect of inertial forces with respect to viscous forces, is typically calculated to identify the type of damping involved in the system. The Reynolds number is estimated as $Re = \rho$ $V_{max}$D/$\mu$, where $\rho$= 1.225 kg/m$^3$ is the density of air, $V_{max}$ is the spider speed, D is the characteristic length and $\mu$ =$1.81 \times 10^{-5}$ kg/(m s) is the viscosity of air. The Reynolds number is calculated for both the spider and the web below. \\

The characteristic length of the spider is estimated to be $D_s \sim\;1.75\;mm$ (fig. 2a) yielding a $Re_s$ $\sim \;$  280. The spider drag is approximated by flow around a sphere at finite Reynolds number as $F_{d,s} = C_sV_s^2$, where $C_s=\frac{1}{2} C_dA_s\rho_{air}$, $C_d$ $\sim$ 1.25 is the drag coefficient at $Re\sim\;$280 and $A_s= \frac{\pi}{4} D_s ^2$ is the characteristic cross-sectional area, and $V_s$ is the spider instantaneous speed~\citep{Vogel2009-gu}. 

Since the silk fibers are orders of magnitude smaller than the spider itself, we expect the Reynolds number for the silk to be much lower. Assuming each silk line as a cylinder that pivots around a stationary substrate so that the web moves at half the speed as the spider ($V_w=V_s/2$), we obtain a Reynolds number $Re_w\sim$ 0.1. For these low  Reynolds numbers, we can estimate the drag on the silk lines using slender body theory in Stoke flows~\citep{Gary_Leal2007-to} as: $F_{dw} = \tfrac{4 \pi L_{w} \mu V_w}{ln{\frac{L_w}{D_w}}}$, where $L_w$ and $D_w$ are the web length and diameter, respectively (see Table 1). Due to the principle of linearity in low Reynolds flow, we can add up the drag contributions from all the radial and capture lines by measuring them and summing up the drag forces. A summary of the values  used in the model are provided in Table 1.



\begin{table}[hbp!]

\label{tab:1} 
\centering
\begin{tabular}{llccc}
                           &                                & \textbf{Simulation 1}         & \textbf{Simulation 2}        &                 \\ \cline{3-4}
\multicolumn{1}{c}{}       & \textbf{Parameter}             & \textbf{Value}                & \textbf{Value}               & \textbf{Source} \\ \hline
\textbf{Spider}            & Mass, $m_s$ (Kg)                  & 1.6$\times 10^{-3}$                    & 1.6$\times 10^{-3}$    & Measured        \\
                           & Characteristic Length, $D_s$ (m)  & 1.3$\times 10^{-3}$                   & 1.3$\times 10^{-3}$    & Measured        \\
                           & Body Length, BL (m)            & 1.75$\times 10^{-3}$                    & 1.75$\times 10^{-3}$    & Measured        \\
                           & Damping coefficient, $C_s$ (Kg/s) & 8.3$\times 10^{-7}$   & 8.3$\times 10^{-7}$   & Calculated     \\ \hline
\textbf{Radial lines}      & Number, $N_r$                    & 8                             & 8                            & Field Data      \\
                           & Young's Modulus, $E_r$ (Pa)       & 0.35$\times 10^{10}$   & 0.45$\times 10^{10}$  & \citep{Gosline1999-jr}   \\
                           & Diameter, $D_r$ (m)               & 1$\times 10^{-6}$      & 1$\times 10^{-6}$     & SEM             \\
                           & Length, $L_r$ (m)                 & 4.5$\times 10^{-2}$    & 4.5$\times 10^{-2}$   & Measured      \\
                           & Equilibrium Length, $L_{r,eq}$ (m)   & 2.34$\times 10^{-2}$   & 2.34$\times 10^{-2}$  & Measured      \\ \hline
\textbf{Tension line} & Young's Modulus, $E_t$ (Pa) & 0.35$\times 10^{10}$ & 0.45$\times 10^{10}$ & \citep{Gosline1999-jr} \\
                           & Diameter, $D_t$ (m)               & 1$\times 10^{-6}$     & 1$\times 10^{-6}$   & SEM             \\
                           & Length, $L_t$ (m)                 & 4$\times 10^{-2}$     & 4$\times 10^{-2}$     & Measured      \\
                           & Equilibrium Length, L$_{t,eq}$ (m)   & 6$\times 10^{-2}$                       & 6$\times 10^{-2}$                       & Measured     \\ \hline
\textbf{Capture lines}     & Number per radial line         & 13                            & 13                           & Measured      \\
                           & Length, $L_c$ (m)                 & 6.5$\times 10^{-3}$      
                           & 6.5$\times 10^{-3}$     & Calculated      \\ \hline
\textbf{Total web lines}   & Total Length, $L_w$ (m)           & 1.03                          & 1.03                         & Calculated      \\
                           & Damping coefficient, $C_w$ (Kg/s) & 2.41$\times$10$^{-4}$    & 2.41$\times$10$^{-4}$  & Calculated      \\ \hline
\textbf{Initial Conditions} & $x_o$ (m)                         & -4.5$\times 10^{-4}$   & -18.5$\times 10^{-4}$  &                 \\
                           & $y_o$ (m)                         & -12$\times 10^{-3}$ & -12$\times 10^{-3}$ &                
\end{tabular}
\caption{\textbf{Simulation parameters and their values.} Summary of the geometrical parameters and physical properties used in the simulation. The source of these parameters is either from literature or measure from field data. The results of simulations 1 and 2 are shown in Fig. 4a and Fig. 5a, respectively.}
\end{table}

\subsection*{2-D Equations of Motion}
Considering all the forces (inertia, web elasticity, and drag) along $x$ and $y$ directions, we write the 2D equations of motion that describe the trajectory of the slingshot spider as a function of time as follows:
\begin{equation}
\begin{split}
\vspace{-2em}
\textbf{x-axis}&: m_s\ddot{x} = -F_{r1x}+F_{r2x}-F_{tx}-F_{dsx}-F_{dwx} \\
\textbf{y-axis}&: m_s\ddot{y} = -F_{r1y} -F_{r2y}-F_{ty}-F_{dsy}-F_{dwy}\\
\end{split}
\end{equation}
\begin{gather*}
\textbf{where:}\\ F_{r1,2x}=\Delta L_{r1,2} \;K_r\;\cos\phi _{r1,2}
\;,\;F_{r1,2y}=\Delta L_{r1,2} \;K_r\;\sin\phi _{r1,2}\\
\;F_{tx}=\Delta L_t\;K_t\;\sin\phi_t\;,\;F_{ty}=\Delta L_t\;K_t\;\cos\phi_t\\
F_{dwx}=C_w\;\frac{\dot{x_s}}{2}\;,\;F_{dwy}=C_w\;\frac{\dot{y_s}}{2}\\
F_{dsx}=C_s\;\ (\dot{x_s}^2 + \dot{y_s}^2) \cos( \arctan{\frac{\dot{y_s}}{\dot{x_s}})}  ,\;F_{dsy}=C_s\;(\dot{x_s}^2 + \dot{y_s}^2) \sin( \arctan{\frac{\dot{y_s}}{\dot{x_s}})}\\
\Delta L_{r1,2}= L_{r1,2} - L_{r,eq}\;,\; L_{r1,2}=\sqrt{L_{r,eq1,2}^2\pm x^2}-L_{r,eq1,2} \\
\sin\phi_{r1,2}=\frac{y}{L_{r1,2}}\;,\;
\cos\phi_{r1,2}=\frac{L_{r,eq1,2}\pm x}{L_{r1,2}}\;,\;\sin\phi_t= \frac{x}{L_t}\\
K_{r1,2}=\frac{E_{rA_r}}{L_{r,eq1,2}}\;,\;
K_{t1,2}=\frac{E_{tA_t}}{L_{t,eq1,2}}\\
\end{gather*}

At $t < 0$, the spider is at static equilibrium at a position ($x_o$,$y_o$) where the radial spring forces ($F_{r1,2}$) are opposed by the force ($F_t$) exerted by the tension line (Fig. 2c1). At $t = 0$, the tension force is set to zero resulting in a net upward motion. This is akin to the spider launching itself when triggered by an external stimulus. The $x$-$y$ trajectories are obtained by numerically solving the equations (1) and (2) using the 4th order Runge-Kutta approach in Matlab. The system starts from static equilibrium at ($x_o$,$y_o$) at $t = 0$. The simulation is stopped after $200\;ms$ has elapsed. 

\section*{Results and Discussion}
\subsection*{Web dynamics in 1D motion}

\begin{figure}
 \includegraphics[width=1\textwidth]{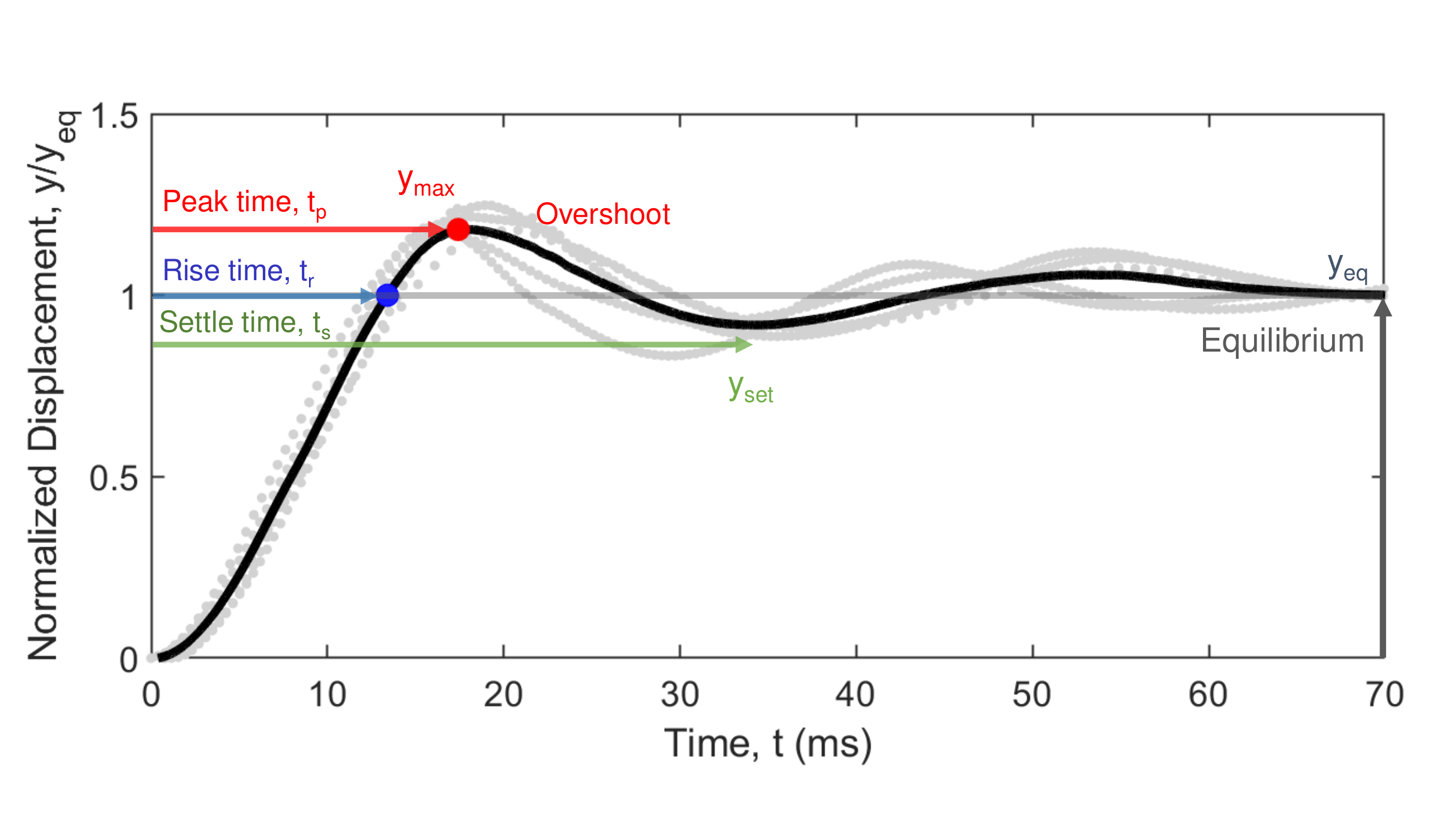}
\caption{\textbf{Defining key parameters of slingshot spider trajectories.} We focus on several spatial and temporal parameter to assess the fit between the mathematical model and the field data. The equilibrium point is set as the final displacement ($y_{eq}$) when the spider comes to a halt. We define the rise time ($t_s$) as the duration from $y=0$ to the first time the spider cross the equilibrium point ($y_{eq}$) and the peak time ($t_p$) is the time taken to reach the maximum displacement ($y_{max}$). We measure the overshoot as the ratio between the max displacement and the equilibrium point ($\frac{y_{max}-y_{eq}}{y_{eq}}$). 
These variables are highlighted here on the average normalized displacements ($n = 5$) in the vertical direction of a single slingshot spider. Finally, we define the settling time as the time it takes the spider to finish one oscillation with an amplitude that falls between $\pm$ one body length of the spider. }
\label{fig:3}      
\end{figure}

To understand the dynamics of the system, we first consider the limiting 1D case, where the motion is purely in the y-direction ($\phi_t=0$ , $x=0 \;, \;\forall t$ ) (Fig. 2c). This assumption may be valid during the initial stages of the motion, where the displacement is almost one-dimensional in the y-direction and relatively negligible in other dimensions (Fig. 1c-d). However, this assumption breaks down during the later stages of the displacement as motion gets more complex. Nevertheless, this approach helps to shed light on several key features of the slingshot motion as well as providing guidance to the iterative exercises of fitting the model to the data. 

We assess the goodness of the fit by using four different spatiotemporal parameters: peak time ($t_p, ms$), settle time ($t_{s}, ms$), overshoot ($\%$), and rise time ($t_r, ms$), which are graphically defined in Fig. 3. The peak time is duration that the spider takes to go from initial position to the maximum displacement $y_{max}$. The overshoot is the percentage offset the spider travels with respect the equilibrium point $\frac{y_{max}-y_{eq}}{y_{eq}}$. The settling time ($t_s$) is the duration that the spider take to go from the initial peak displacement to the first oscillatory peak that falls between $\pm$  on the body length (BL) above the equilibrium point. We assume that at this point, the slingshot motion has practically come to a halt and the spider is able to reset the elastic loading.

The rise time $t_s$ corresponds to the time the spider takes to go from $y = y_o$ to $y = y_{eq}$ during the first oscillation (Fig. 3). During this time $t<t_s$, the spatiotemporal dynamics are mainly governed by the radial line forces $F_r$ along with drag $F_d$, while the tension line force $F_t$ is zero ($\Delta L_t < 0$). The initial rise time is $t_s=2\pi$/$\omega_{d,n}$ where $\omega_{d,n}$ is the damped natural frequency.  Once $\Delta L_t \ge 0$, the tension line becomes engaged and the restoring force $F_t$ starts influencing the dynamics of the system. Specifically, the damped natural frequency becomes proportional to the combined effect of both the radial lines and the tension line ( $\omega_n \propto (K_r+K_t)^\frac{1}{2}$ when $y>y_{eq}$ ).

The mass subsequently oscillates around the equilibrium point with diminishing amplitudes before eventually stopping as the kinetic and potential energies get dissipated by drag. An interesting geometrical feature of the model is that as the mass approaches the equilibrium point ($y \rightarrow  0 $, $\phi \rightarrow 0 $), the projected radial forces in the y direction drop sharply.



\subsection*{Slingshot Dynamics in a 2D motion}

Next, we solve the equations of motion in 2D while changing the physical and geometrical parameters to better fit the displacement of the spider. 
To avoid major variations in geometry and initial conditions, we validate the simulation outputs with multiple firing events ($n = 5$) of a single spider (Fig. 1b-d). By matching the initial conditions of the model $(x_o,y_o)$ to the experimental data, we show the output of the model in Fig. 4a, referred to as Simulation 1. The model captures the rise time, peak time and maximum peak with less than $5\%$ error as quantified in Table 2. However, the model does not accurately capture the secondary oscillations ($t>30$ ms), and also predicts a $\sim\;20\;ms$ longer setting time compared to experimental observations. In the x-direction, the model captures both the amplitudes and frequency of the oscillations with a phase shift, as seen qualitatively with the $x$-$y$ map (Fig. 4d).

The temporal evolution of the underlying forces is also computed and highlighted in conjunction with the y-displacement. The radial forces projected into the y-direction start with a combined maximum value of around $\sim 7.5\times$10$^{-4}$ N or $75$ dynes. Interestingly, this falls within the force ranges previously measured in the radial silk lines of slingshot spider webs using a custom-built tension apparatus by Coddington~\citep{Coddington1986-du}. The net radial forces in the $x$-direction are relatively smaller and become more significant at $t>30\;ms$ as the spider approaches equilibrium. The tension line force $F_t$ is only enabled when $\Delta L_t > L_{t,eq}$, and is observed as a sharp increase in the negative direction (retarding spider motion) around $t=\;14\;ms$ (Fig. 4b). The drag forces in the $y$-direction start from zero and rapidly increase in magnitude as the spider approaches $V_{s,max}$ before decreasing as the mass comes to a halt. We note that the radial web forces and drag forces are always acting in opposite directions - the web forces driving motion and drag damping it.  

\begin{figure}
 \includegraphics[width=1\textwidth]{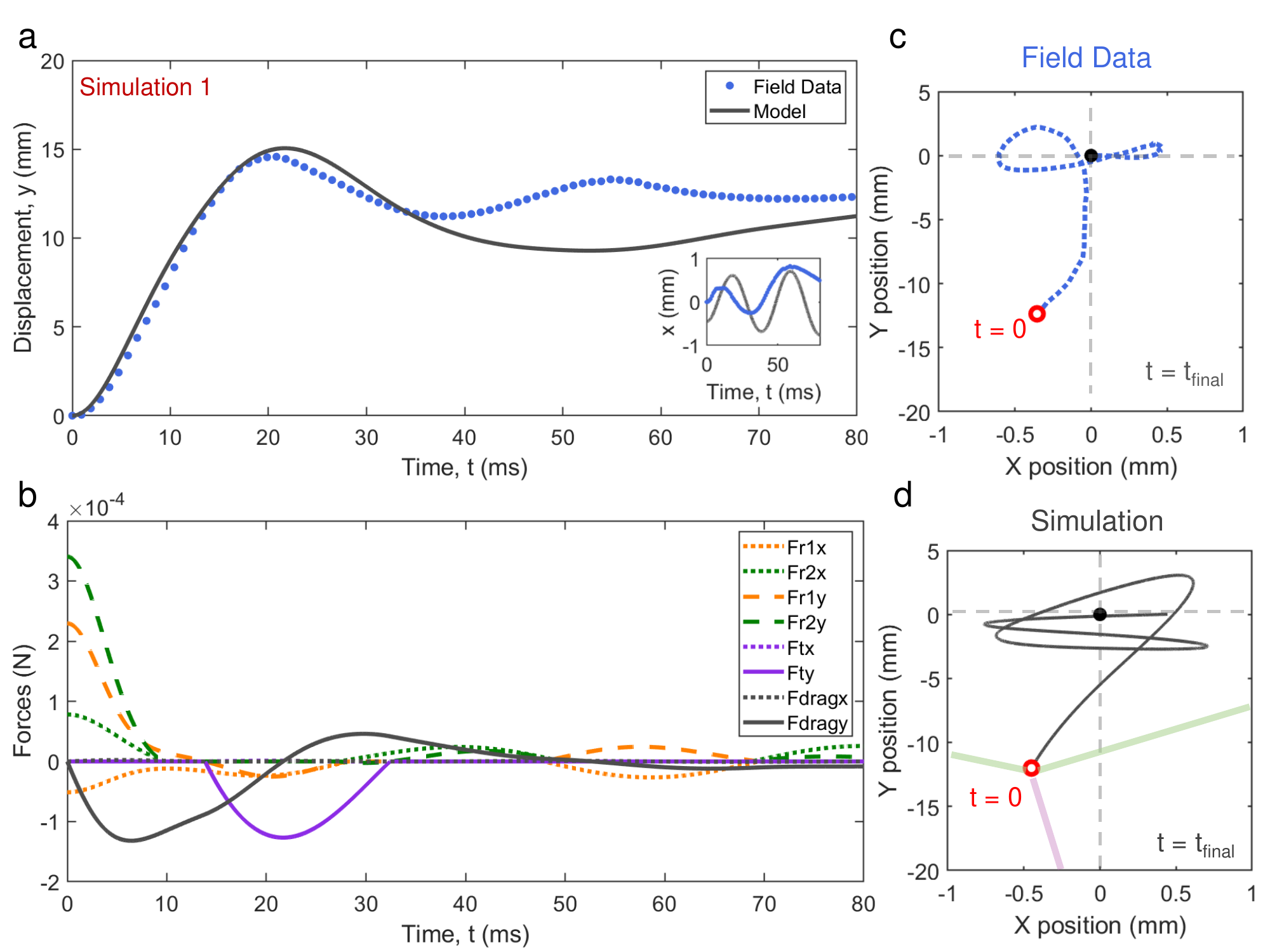}
\caption{ \textbf{Simulation 1: Slingshot spider dynamics in 2-D.} \textbf{a} We obtain the $x$-$y$ displacement of the spider by solving the equations of motion (Eq. 1). The simulation output matches the rise time, maximum peak and maximum output $y_{max}$, as well the minimal oscillations in the $x$ direction. However, beyond the first fluctuation, the model does not predict any subsequent oscillations in the $y$-direction. \textbf{b} The temporal evolution of the forces reveals that the system starts with forces in the radial lines in the $x$-$y$ directions. As the system picks up speed, drag from the web resists the motion - drag from spider is negligible. The restoring force in the tension line which has been zero so far, steps in as the tension line becomes stretched ($t\sim15\;ms$). \textbf{c,d } The simulation also predicts the apparent  2D displacement of the spider starting from from ($x_o,y_o$) at $t=0$. The simulation show a qualitatively similar trajectory in the $x$-$y$ plane, as the spider initially catapults in the $y$-direction before undergoing low amplitude oscillations in the $x$-direction as it slowly approaches equilibrium.  }
\label{fig:4}      
\end{figure}

Two underlying assumptions made in this model are that the trajectory of the slingshot spider is confined in a 2D plane and that the spider behaves as a point mass object. In reality, the slingshot motion is a more complex 3D motion, but we only record a projection of the motion in 2D with a single high-speed camera in the field. Moreover, the mass of the spider is not equally distributed (asymmetric body) with most of its mass not facing the tension line. This unequal distribution of mass causes the spider to behave like a 3D inverted pendulum that can rotate around the point of intersection between the spider and the silk (as seen in SI Movie 1)~\citep{Han2019ExternalWeb.}. The model also ignores any displacement biases that might arise due to asymmetries in the web structure, since the spider constructs the web in small plant branches. To highlight the consequences of displacement biases in an orthogonal direction ($z$ direction), we consider an exaggerated case in the initial value of $x$ where we run the simulation after multiplying $x_0$ by 4. The results shown in Fig. 4 show better agreement between the model and field data, and the secondary oscillations in the experimental data are matched well by the model, including matching of the settling time (Table 2).



\begin{figure}
\centering
 \includegraphics[width=0.9\textwidth]{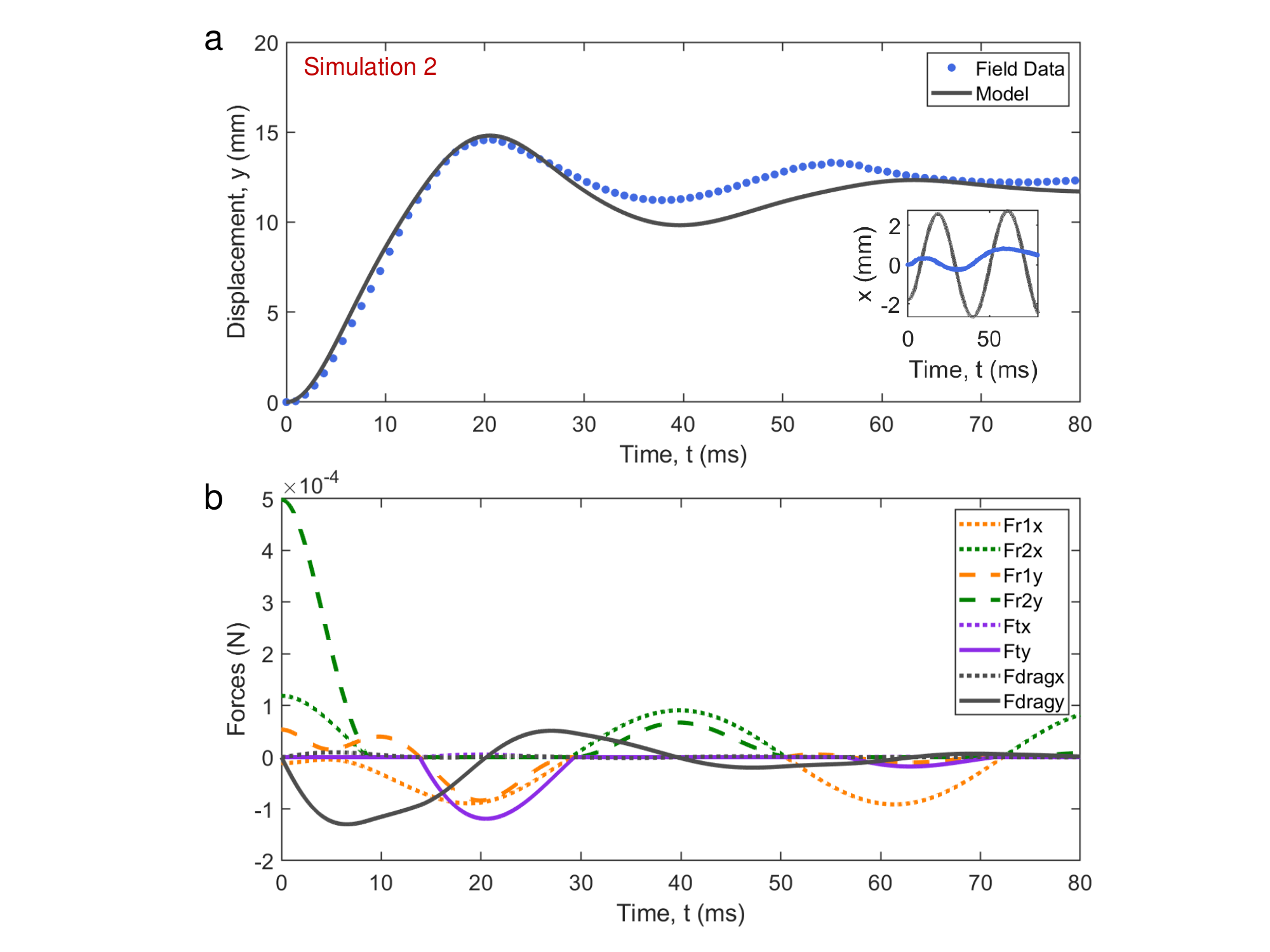}
\caption{ \textbf{Simulation 2 with larger contributions from $x$-motion.} \textbf{a} To capture the dynamics of the subsequent vibrations in the $y$-direction, we run a second simulation with an exaggerated initial displacement in the x-direction (4 $\times$ the one used in simulation 1). This highlights the additional layer of complexity imposed by an amplified orthogonal motion. The model matches the field data better in terms of the temporal parameters as well as the subsequent low amplitude fluctuations after the first overshoot. In reality, the spider motion is  3D, which may include twisting and rotation due to asymmetries in mass and geometry (See SI Movie 1). These extraneous influences become more significant at low amplitudes and frequencies. \textbf{b} The evolution of the forces in the $x-y$ plane show similar dynamics as seen earlier in Fig. 4b. Interestingly, the tension force in the $x$-direction becomes more pronounced here as the influence in the $x$-direction is amplified due to the initial conditions.}
\label{fig:5}      
\end{figure}

\begin{table}[]
\centering
\begin{tabular}{lccccc}
\textbf{Parameter}       & \textbf{Field data} & \textbf{Simulation 1} & \textbf{Rel. error \%} & \textbf{Simulation 2} & \textbf{Rel. error \%} \\ \hline
Rise time, $t_r$         & 9.48 ms             & 9.6 ms               & 1.25\%                 & 9.6 ms               & 1.25\%                 \\
Peak, $y_{max}$          & 14.58 mm            & 15.07 mm              & 3.36\%                 & 14.81 mm              & 1.57\%                  \\
Peak time, $t_p$         & 20.81 ms            & 20.25 ms              & 2.76\%                 & 20.5 ms              & 1.53\%                 \\
Overshoot \%             & 17.93 \%            & 30.5 \%              & 12.57\%                    & 23.89 \%              & 5.96\%                 \\
Settling time, $t_{s}$ & 37.84 ms            & 60 ms              & 58.5\%                & 40 ms              & 5.71\%                
\end{tabular}

\caption{\textbf{Comparison of simulation to single representative field data point} Summary of the spatiotemporal parameter outputs of simulations 1 and 2.}
\end{table}

In summary, despite the major dimensional reductions assumed in developing this simplified mathematical model, our model still captures salient spatiotemporal dynamic features that arise due to the complex design of the slingshot spider web. The values used in the model fall within the biological boundaries in terms of physical properties and geometrical constraints (Tables 1,2). Next we explore the rich parameter space provided by this model, by evaluating the web dynamics under the influence of changing web stiffness $K_{t}$, tension line stiffness $K_{r}$, and drag force coefficient $C_{w}$.

\subsection*{Slingshot spider dynamics as a function of web parameters ($K_t, K_r, C_d$) }

Here we explore the effect of three  parameters - radial stiffness $K_r$, tension line stiffness $K_t$ and web drag $C_w$ -  on the dynamics of the slingshot motion. These parameters capture most of the mechanical and geometrical characteristics of the web and dictate the significant  dynamics of the system. Simulations are performed by changing one of these parameters while keeping the others constant. For a convenient physical interpretation, we normalize the output displacements by the spider body length (BL) while using the results and parameters ($K_{t,ss}$, $K_{r,ss}$ and $C_{w,ss}$) from Simulation 2 as a reference. We showcase the resulting normalized displacement curves with respect to the reference plot in Fig. 6 and compare the sensitivity of the model outputs in terms of the aforementioned spatiotemporal parameters in Fig. 7.

\begin{figure} [htb]
 \includegraphics[width=1.1\textwidth]{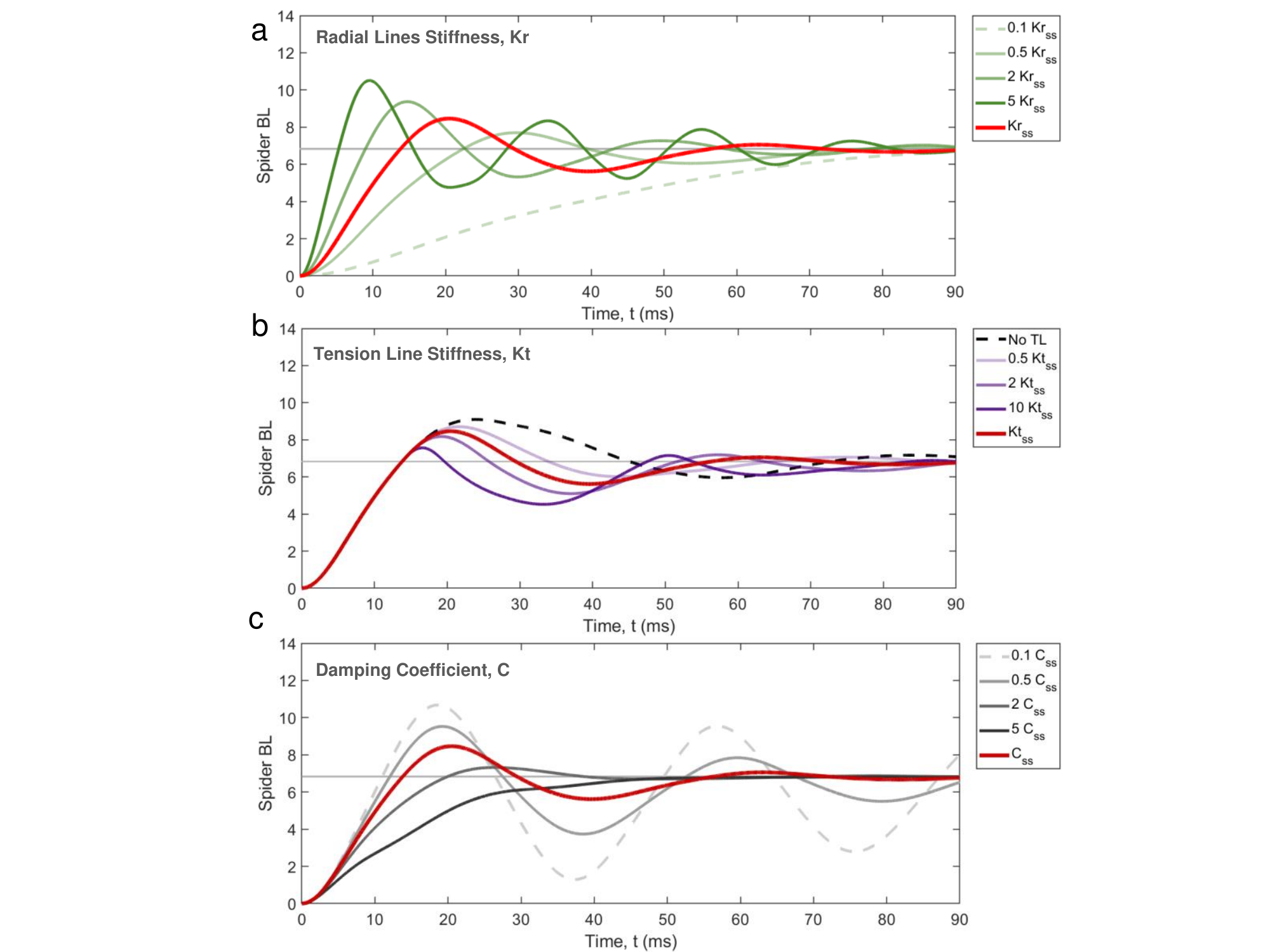}
\caption{\textbf{Exploring web dynamics as a function of model parameters.} We explore the impact of the different forces on the slingshot dynamics obtained from Simulation 2. The output displacements are normalized by the spider body length (BL) and parameters ($K_{r,ss}$, $K_{r,ss}$ and $C_{w,ss}$) from Simulation 2 (Fig. 5a) as a reference. \textbf{a} Increasing the radial lines stiffness $K_r$ leads to both a faster increase in the initial displacement as well as higher overshoots ($\sim$4 BL above equilibrium) with subsequent increasingly large amplitude vibrations around the equilibrium point. An order of magnitude decrease in the radial lines stiffness causes the system to slowly approach the equilibrium point without any observable overshoot. \textbf{b} Changing the tension line stiffness $K_t$ only affects the dynamics past the equilibrium line. Without any tension force the system achieves a higher overshoot and more subsequent vibrations. As the tension line stiffness increases, the vibrations peaks become sharper above the equilibrium with larger and slower undershoot. \textbf{c.} The model reveals high sensitivity to web drag $C_w$. Halving the damping coefficient causes the system to significantly vibrate over two body lengths around the equilibrium line. Doubling the damping coefficient leads to a small overshoot as the system quickly approaches equilibrium without any noticeable oscillations, similar to critically damped oscillator. }
\label{fig:6}      
\end{figure}

Fig. 6a shows the normalized displacement curves when the radial line's stiffness $K_r$ is changed while everything else remains the same.  We observe that, for an order of magnitude decrease in the value of $K_r$, the web loses it's ability to oscillate and slowly asymptotes to the equilibrium point. This is similar to an overdamped system where drag dissipation dominates. Increasing $K_{r,ss}$ causes a faster rise in the displacement as well as a larger overshoot. These results are quantified in Fig. 7 which shows that as the radial stiffness increases, the spider travels a longer distance in a shorter time duration and takes a shorter time to settle. We (and others~\citep{Eberhard1986-kw,Coddington1986-du}) have observed that slingshot spiders fire their webs in response to external vibratory/sound clues (finger snapping). It is therefore possible that spiders could release their webs in response to the nearby frequency of an insect's wing beat. Furthermore, it is key that the spider resets its web rapidly, as to not miss out on potential prey. Our model reveals that the slingshot $K_r$ appears to balance  finite overshoot at smaller peak and settle times, which could facilitate reaching a flying prey at a distance quickly, with minimizing oscillations to reset, and repeat motion if unsuccessful.

From the biomechanics context of the slingshot spiders motion, what does it mean to change parameters such as the stiffness of the radial lines? Stiffness depends on intrinsic molecular and geometric properties and is defined as $K_r=\frac{N_r E_r A_r}{L_{r,eq}}$. These properties vary among different species of spiders, and have never been measured specifically for \textit{Epeirotypus} sp.. However, the young's modulus ranges between 1-10$\times 10^9$ Pa in the literature, depending upon external mechanical conditions such as such as strain, strain rate and environmental conditions such as temperature and humidity~\citep{Su2016-ap, Yazawa2020-ro, Agnarsson2009-pt}. Another strategy to modify the radial stiffness would be changing by other geometrical parameters such as the equilibrium length $L_{r,eq}$ or the area $A_r$. For example Eberhard reports direct observations of \textit{Epeirotypus} sp. using its legs to adjust the tension in the radial and sticky lines (by effectively changing $L_{r,eq}$) during web construction~\citep{Eberhard1981-as}.

Next, we examine the role of the tension line highlighted in Fig. 6b. Since the tension line is not originally stretched ($\Delta L_t < L_{t,eq}$), varying the tension line stiffness does not affect the rise time $t_s\approx 10$ ms. Beyond this point, the tension line is engaged and applies a pulling force on the spider. In the limit of no tension line ($K_t\;=0$), the system overshoots to almost two spider BL above the equilibrium point before slowly oscillating back to equilibrium. As the stiffness of tension increases, the overshoot decreases while the undershoot increases. For instance, at $10\;K_{t,ss}$, the maximum displacement decreases to less than one body length above the equilibrium before bouncing into almost two body lengths below the equilibrium point. This is due to the fact that at slightly above equilibrium, the time scale is mostly determined by the properties of the tension line, namely $t =\frac{2\pi}{\omega _{nd}} \sim \sqrt{\frac{m_s}{K_t}}$. Fig. 7b summarizes these results showing a monotonic decrease from 30$\%$ to almost 8$\%$ across four order of magnitudes in $K_t$. Meanwhile, the peak time is not as strongly influenced, decreasing by only $\sim10\;ms$ over four order span in $K_t$. Interestingly, we observe a minimum of around $40\;ms$ near the reference value in the settling time due to the increase in the undershoot as $K_t$ increases.

From a biomechanics context, our model suggests that, besides allowing the spider to load elastic energy and keep the web in static equilibrium, the tension line also plays an integral role during the slingshot motion. Without a tension line, the slingshot spider is completely at the mercy of air drag near the equilibrium point as radial forces become negligible due to geometry (see SI text Fig. 2). The tension line stiffness assists in preventing too much overshoot while decreasing the settling time. Controlling the overshoot and settling the web faster may allow the spider to reach the insect in its web faster. It may also reduce the time to reset and repeat its hunting motion in case of a missed prey capture. The tension line also allows the spider to control its vertical displacement at intermediate tension line lengths actively as observed in field experiments (See SI Movie 2).


Finally, we look at the effect of the web drag on the slingshot dynamics. Compared to the other two parameters, we observe that the model is highly sensitive to the drag coefficient.At a high damping coefficient (2 to 10$\times C_{w,ss}$),  the system goes rapidly to an overdamped regime whereby the system gradually approaches equilibrium without any oscillations. At lower damping coefficient (0.1 to 0.5$\times C_{w,ss}$),  the system becomes  highly underdamped and vibrates with several body lengths around the equilibrium point. This is further highlighted in Fig. 7c, which shows that overshoot may reach up to 60$\%$ at 0.01 $C_{w,ss}$. At values larger than 10 $C_{w,ss}$, the overshoot is minimal or even non existent. The high sensitivity of the model to damping may be further exemplified in the settling time which shows a minimum near the reference value $C_{w,ss}$. Above the minimum value, the system is overdamped and takes a long time to reach the $\pm$ 1BL envelope around equilibrium. Below that minimum, the system becomes highly underdamped and fluctuates with high amplitudes around equilibrium.


\begin{figure}
 \includegraphics[width=1\textwidth]{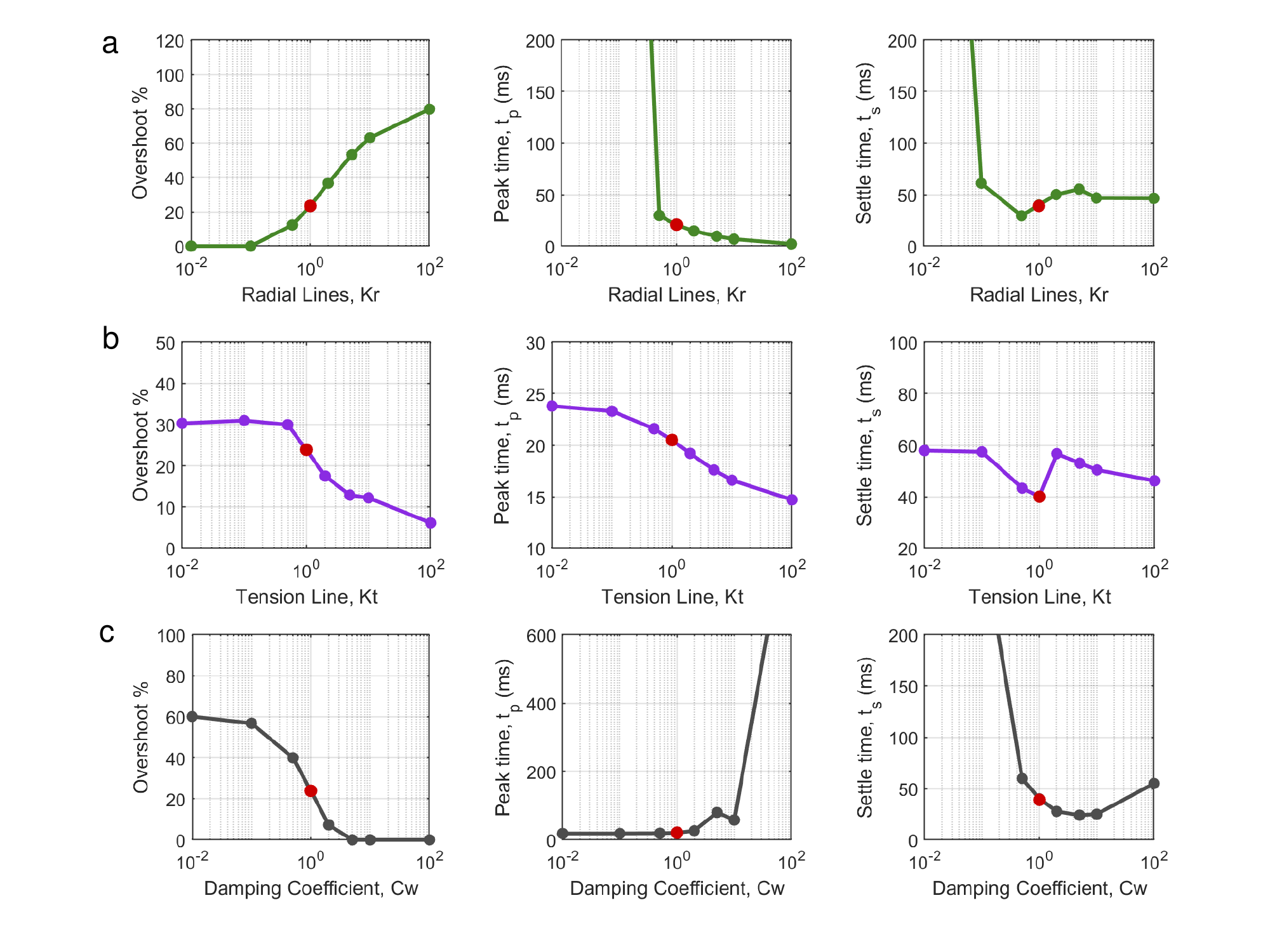}
\caption{\textbf{Summary of web dynamics as a function of model parameters.} We quantify the changes of the various simulations by focusing on the percentage overshoot, peak time ($t_{p}$) and settling time ($t_{s}$). We highlight the reference values in red \textbf{a.} The overshoot increases monotonically from 0 (no peak) to 80 $\%$ with the increase in $K_{r,ss}$. At low $K_{r,ss}$, the system does not vibrate around the equilibrium point causing $t_p$ and $t_s$ to go beyond the max simulation time of 200 ms. This behavior is tantamount to an overdamped dynamical system as the damping forces overcome the spring forces causing the . \textbf{ b.} The overshoot decreases by almost 35$\%$ as  $K_t$ increases from $0.01$ to $100 \times K_t$. Similarly, the peak time $t_p$ slightly decreases from $24\;ms$ to $15\;ms$ as $K_t$ increases. Interestingly, the settling time ($t_s$) decreases to a minimum of $40\;ms$ around the reference value of $K_t$ before sharply increasing. This is mainly due to the larger undershoot causes by a higher $K_t$ \textbf{c.} The simulation is highly sensitive to the damping coefficient. The overshoot decreases from around 60$\%$ to almost no overshoot as damping increases. The settling time shows a minimum around the reference point as low drag causes the system to oscillate violently around the equilibrium point while a high drag causes it to slowly approach the equilibrium. Both effects yield an increase in the settling time.  }
\label{fig:7}      
\end{figure}

 \subsection*{Limitations of the model and future work}

Our work aims to provide a reduced-order approach to modeling the complex dynamics in slingshot spider webs. However, we make certain assumptions in our model that can be further improved in future work. 
\subsubsection*{Dissipation due to aerodynamic drag and molecular friction}
Our model shows that aerodynamic drag at an intermediate to low Reynolds number adequately describes the damping experienced by the spider and its web. The model ignores the possible contribution of viscous dissipation within the viscoelastic spider threads.
The relative effect of these two sources of dissipation in oscillating spider webs is underexplored and still an active area of research. Past work highlighted the importance of aerodynamic drag in the context of  “ballooning spiders” \citep{Sheldon2017,Suter1992} and the stopping of flying insects by orb weavers webs \citep{Sensenig2012-id,Kelly2011-lc}. Other work however completely dismissed aerodynamic drag and attributed dissipation primarily to internal viscous damping \citep{Ko2004,Ayon}. A summary of studies exploring the role of dissipation, internal and external, is provided in SI Table 1. 

Owing to the fineness of spider silk having a diameter $ <  4\;\mu m$ \citep{Ko2004}, the characterization of the mechanical properties of the spider silk had been mainly experimentally limited to tensiometry or impulsive loading at lower strain rates. Replicating the high strain rates faced by the webs of the slingshot spiders studied in this work ($> 60\;s^{-1}$) is experimentally challenging. In addition, the mechanical properties and structural geometries of spiders’ web silk are highly dependent on environmental conditions such as temperature, humidity, wind conditions, water content (supercontractility) and surrounding conditions such as the location and flexibility of the substrate \citep{Gosline1999-jr}. This poses further difficulties in replicating the native conditions of spiders to experimentally examine the mechanical properties of the spider silk. Overall, our model reveals that viscous damping (in silk) is not necessary to capture the underdamped behaviour in the slingshot spider, but it could still be important and will be focus of future work (See SI text fig. 3).

\subsubsection*{Geometrical constraints}
Our simulations also assume no coupling between the parameters. For instance, orb webs vary tremendously across species in the numbers of radii and rows of capture spirals~\citep{Craig1987, Sensenig2010}), but slingshot spider webs are relatively similar to one another in having relatively few radii and around ten rows of capture silk in  their webs~\citep{Eberhard1986-kw}. This might suggest geometric constraints on slingshot spider web topology and dimensions. Thus, this model may be extended to examine other web topologies and silk dimensions to test the hypothesis that the evolution of theridiosomatid web architecture is limited in part by optimizing the amount of aerial damping during the slingshot motion

\section*{Conclusion}


We develop a 2D mathematical model to simulate the dynamics of the slingshot spider powered by its conical web geometry and tension line. We validate the model with experimental results and explore the sensitivity of various physical parameters governing web forces. We find that web parameters are finely-tuned to yield an underdamped oscillating web, that enables the spider to displace finite distances quickly, while minimizing residual oscillations in the web. These design parameters may enable the spider to exploit a risky hunting strategy of catching flying insects in mid-air while minimizing oscillations in its web due to its rapid movements. At the same time, the spider should be able to sense the vibrations induced by the flying preys and discern them from possibly faulty ones induced by the surrounding ambient air ~\citep{Craig1985-oe, Eberhard1981-as}. Though we have presented a first approach for understanding this fascinating slingshot spider dynamics, open questions remain about the  molecular  structure  of the radial silk and tension line, as well as their  3D-web topology that enables their agile power packed performance. These research questions open up rich avenues for multidisciplinary research,  while furthering our  knowledge  and appreciation  of  arachnids  and  their  ingenious  engineering strategies for locomotion and survival.

\section*{Declarations}

\subsection*{Funding} 
S.J. acknowledges funding support from the NSF under grant no. CBET-2002714. M.S.B acknowledges funding support through NSF award number 1817334 and CAREER 1941933 and National Geographic Foundation (NGS-57996R-19). T.A.B. acknowledges funding support from the NSF (IOS-1656645)

\subsection*{Conflicts of interest} 
The authors declare no competing interests.

\subsection*{Availability of data and material} 
All Matlab codes and data for this article are accessible here: \url{https://github.com/bhamla-lab/slingshotspider2021}

\subsection*{Ethics approval} 
All applicable international, national, and institutional
guidelines for the care and use of animals were followed. 

\subsection*{Author's contribution} 
SA, MSB collected field data. EJC, SA, SJ developed mathematical model. EJC, SA and SH analyzed data. EJC conducted simulations. All authors contributed to editing, interpreting and writing the manuscript. TAB, SJ and MSB managed funding and resources.  

\begin{acknowledgements}
We thank Jaime Navarro for his excellent field guide services in the Peruvian Amazon Rainforest. 
\end{acknowledgements}

%

\bibliographystyle{spbasic}      
\bibliography{references_SB,references}

%
%

\end{document}